\documentclass{elsart1p} 
\usepackage{epsfig} 
\usepackage{amssymb} 

\journal{Physics Letters B} 

\begin{document} 

\begin{frontmatter} 

\title{New results on Mesonic Weak Decay of $p$-shell 
$\Lambda$-Hypernuclei} 
\author{The FINUDA Collaboration} 
\linebreak 
\author[a,b]{M.~Agnello}, \author[c]{A.~Andronenkov}, \author[d]{G.~Beer}, 
\author[e]{L.~Benussi}, \author[e]{M.~Bertani}, \author[f]{H.C.~Bhang}, 
\author[g,h]{G.~Bonomi}, \author[i,b]{E.~Botta \thanksref{cor1}}, 
\author[j,k]{M.~Bregant}, \author[i,b]{T.~Bressani}, \author[i,b]{S.~Bufalino}, 
\author[l,b]{L.~Busso}, \author[b]{D.~Calvo}, \author[j,k]{P.~Camerini}, 
\author[n,c]{B.~Dalena}, \author[i,b]{F.~De Mori}, \author[n,c]{G.~D'Erasmo}, 
\author[e]{F.L.~Fabbri}, \author[b]{A.~Feliciello}, \author[b]{A.~Filippi}, 
\author[n,c]{E.M.~Fiore}, \author[h]{A.~Fontana}, \author[u]{H.~Fujioka}, 
\author[h]{P.~Genova}, \author[e]{P.~Gianotti}, \author[k]{N.~Grion}, 
\author[e]{O.~Hartmann}, \author[f]{B.~Kang}, \author[n]{V.~Lenti}, 
\author[e]{V.~Lucherini}, \author[i,b]{S.~Marcello}, \author[x]{T.~Maruta}, 
\author[r]{N.~Mirfakhrai}, \author[s,h]{P.~Montagna}, \author[t,b]{O.~Morra}, 
\author[u]{T.~Nagae}, \author[p]{D.~Nakajima}, \author[v]{H.~Outa}, 
\author[e]{E.~Pace}, \author[c]{M.~Palomba}, \author[c]{A.~Pantaleo}, 
\author[h]{A.~Panzarasa}, \author[c]{V.~Paticchio}, \author[k]{S.~Piano}, 
\author[e]{F.~Pompili}, \author[j,k]{R.~Rui}, \author[z]{A.~Sanchez Lorente}, 
\author[w]{M.~Sekimoto}, \author[n,c]{G.~Simonetti}, \author[w]{A.~Toyoda}, 
\author[b]{R.~Wheadon}, \author[g,h]{A.~Zenoni} 
\linebreak 
and 
\linebreak 
\author[zz]{A.~Gal} 
\address[a]{Dipartimento di Fisica, Politecnico di Torino, Corso Duca degli 
Abruzzi 24, Torino, Italy} 
\address[b]{INFN Sezione di Torino, via P. Giuria 1, Torino, Italy} 
\address[c]{INFN Sezione di Bari, via Amendola 173, Bari, Italy} 
\address[d]{University of Victoria, Finnerty Rd., Victoria, Canada} 
\address[e]{Laboratori Nazionali di Frascati dell'INFN, via. E. Fermi, 40, 
Frascati, Italy} 
\address[f]{Department of Physics, Seoul National University, 151-742 Seoul, 
South Korea} 
\address[g]{Dipartimento di Meccanica, Universita' di Brescia, via Valotti 9, 
Brescia, Italy} 
\address[h]{INFN Sezione di Pavia, via Bassi 6, Pavia, Italy} 
\address[i]{Dipartimento di Fisica Sperimentale, Universita' di Torino, 
Via P. Giuria 1, Torino, Italy} 
\address[j]{Dipartimento di Fisica, Universita' di Trieste, via Valerio 2, 
Trieste, Italy} 
\address[k]{INFN Sezione di Trieste, via Valerio 2, Trieste, Italy} 
\address[l]{Dipartimento di Fisica Generale, Universita' di Torino, 
Via P. Giuria 1, Torino, Italy} 
\address[n]{Dipartimento di Fisica Universita' di Bari, via Amendola 173, 
Bari, Italy} 
\address[u]{Department of Physics, Kyoto University, Sakyo-ku, Kyoto Japan} 
\address[x]{Department of Physics, Tohoku University, Sendai 980-8578, Japan} 
\address[r]{Department of Physics, Shahid Behesty University, 19834 Teheran, 
Iran} 
\address[s]{Dipartimento di Fisica Teorica e Nucleare, Universita' di Pavia, 
via Bassi 6, Pavia, Italy} 
\address[t]{INAF-IFSI, Sezione di Torino, Corso Fiume 4, Torino, Italy} 
\address[p]{Department of Physics, University of Tokyo, Bunkyo, Tokyo 
113-0033, Japan} 
\address[v]{RIKEN, Wako, Saitama 351-0198, Japan} 
\address[z]{Instit\"{u}t fur Kernphysik, Johannes Gutenberg-Universit\"{a}t 
Mainz, Germany} 
\address[w]{High Energy Accelerator Research Organization (KEK), Tsukuba, 
Ibaraki 305-0801, Japan} 
\address[zz]{Racah Institute of Physics, The Hebrew University, Jerusalem 
91904, Israel} 
\thanks[cor1]{Corresponding author. Fax: +39~011~6707324; e-mail address: 
botta@to.infn.it} 

\begin{abstract} 
The FINUDA experiment performed a systematic study of the charged mesonic 
weak decay channel of $p$-shell $\Lambda$-hypernuclei. Negatively charged 
pion spectra from mesonic decay were measured with magnetic analysis for the 
first time for ${^{7}_{\Lambda}\mathrm{Li}}$, ${^{9}_{\Lambda}\mathrm{Be}}$, 
${^{11}_{\Lambda}\mathrm{B}}$ and ${^{15}_{\Lambda}\mathrm{N}}$. 
The shape of the $\pi^{-}$ spectra was interpreted 
through a comparison with pion distorted wave calculations that take into 
account the structure of both hypernucleus and daughter nucleus. Branching 
ratios $\Gamma_{\pi^{-}}/\Gamma_{tot}$ were derived from the measured spectra 
and converted to $\pi^{-}$ decay rates $\Gamma_{\pi^{-}}$ by means of known 
or extrapolated total decay widths $\Gamma_{tot}$ of $p$-shell 
$\Lambda$-hypernuclei. Based on these measurements, the spin-parity assignment 
$1/2^+$ for ${^{7}_{\Lambda}\mathrm{Li}}$ and 
$5/2^+$ for ${^{11}_{\Lambda}\mathrm{B}}$ ground-state are confirmed and 
a spin-parity $3/2^+$ for ${^{15}_{\Lambda}\mathrm{N}}$ ground-state is 
assigned for the first time. 
\end{abstract} 


\end{frontmatter} 

\section{Introduction} 
\label{intro} 

A $\Lambda$-hypernucleus in its ground-state decays to non-strange nuclear 
systems through the mesonic (MWD) and non-mesonic (NMWD) weak decay 
mechanisms. In MWD the $\Lambda$ hyperon decays to a nucleon and a pion 
in the nuclear medium, similarly to the weak decay mode in free space: 
\begin{eqnarray} 
\Lambda_{free} & \rightarrow & p + \pi^{-} + \mathrm{37.8~MeV}\ \ \ (64.2 \%) 
\\   & & n + \pi^{0} + \mathrm{41.1~MeV}\ \ \ (35.8 \%) 
\label{lfree} 
\end{eqnarray} 
in which the emitted nucleon (pion) carries a momentum $q \approx 100$ MeV/c. 
For a $\Lambda$-hypernucleus, the total decay width (or equivalently the 
decay rate) $\Gamma_{tot}({^A_{\Lambda}\mathrm{Z}})$ is given by the sum 
of the mesonic decay width ($\Gamma_{m}$) and the non-mesonic decay width 
($\Gamma_{nm}$), where the first term can be further expressed as the sum 
of the decay widths for the emission of negative ($\Gamma_{\pi^{-}}$) and 
neutral ($\Gamma_{\pi^{0}}$) pions: 
\begin{equation} 
\Gamma_{tot}(^A_{\Lambda}\mathrm{Z}) = \Gamma_{\pi^{-}} + \Gamma_{\pi^{0}} 
+ \Gamma_{nm}, 
\label{gamma} 
\end{equation} 
with $\Gamma_{tot}(^A_{\Lambda}\mathrm{Z})$ expressed in terms of the 
hypernuclear lifetime as: 
\begin{equation} 
\Gamma_{tot}(^A_{\Lambda}\mathrm{Z}) = {\hbar}/ \tau(^A_{\Lambda}Z). 
\label{tau} 
\end{equation} 
MWD is suppressed in hypernuclei with respect to the free-space decay due 
to the Pauli principle, since the momentum of the emitted nucleon is by far 
smaller than the nuclear Fermi momentum ($k_F \simeq 270$ MeV/c) in all 
nuclei except for the lightest, $s$-shell ones. 

The theory of hypernuclear MWD was initiated by Dalitz \cite{dal1,dal2}, 
based on a phenomenological Lagrangian describing the elementary decay 
processes (1) and (2), and motivated by the observation of MWD reactions in 
the pioneering hypernuclear physics experiments with photographic emulsions 
that provided means of extracting hypernuclear ground-state spins and 
parities; see Ref.~\cite{davis} for a recent summary. 
Following the development of counter techniques for use in $(K^-,\pi^-)$ 
and $(\pi^+,K^+)$ reactions in the 1970s and 1980s, a considerable body of 
experimental data on $\Gamma_{\pi^{-}}$ and/or $\Gamma_{\pi^{0}}$ is now 
available on light $\Lambda$-hypernuclei up to ${^{12}_{\Lambda}\mathrm{C}}$: 
${^{4}_{\Lambda}\mathrm{H}}$ \cite{outa1}, ${^{4}_{\Lambda}\mathrm{He}}$ 
\cite{outa2}, 
${^{5}_{\Lambda}\mathrm{He}}$ \cite{szym}, ${^{11}_{\Lambda}\mathrm{B}}$ and 
${^{12}_{\Lambda}\mathrm{C}}$ 
\cite{szym, saka, noumi,sato}. 

Comprehensive calculations of the main physical entities of MWD were performed 
during the 1980s and 1990s for very light $s$-shell \cite{motoba4,motoba2}, 
$p$-shell \cite{motoba2,motoba3,motoba1} and $sd$-shell hypernuclei 
\cite{motoba2,motoba1}. The basic ingredients of the calculations are the 
Pauli suppression effect, the enhancement of MWD owing to the pion-nuclear 
polarization effect in the nuclear medium as predicted for MWD in 
Refs.~\cite{bando,oset}, the sensitive final-state shell-structure dependence, 
and the resulting charge dependence of the decay rates. 

An important ingredient of MWD calculations is the choice of pion-nucleus 
potential which generates pion-nuclear distorted waves that strongly affect 
the magnitude of the pionic decay rates. Indeed, for low-energy pions, 
the pion-nucleus potential has been studied so far through $\pi$-nucleus 
scattering experiments \cite{friedman1} and measurements of $X$-rays from 
pionic atoms \cite{friedman2}; the study of MWD in which a pion is created 
by the decay of a $\Lambda$ hyperon deep inside the nucleus offers important 
opportunities to investigate in-medium pions and to discriminate between 
different off-shell extrapolations inherent in potential models. For this 
reason MWD continues to be an interesting item of hypernuclear physics, 
and precise and systematic determinations of $\Gamma_{\pi^{-}}$ and 
$\Gamma_{\pi^{0}}$ are very welcome. 

In the present work we report on new measurements by the FINUDA experiment of 
MWD of hypernuclei in the $p$-shell, comparing the measured $\pi^-$ spectra 
and decay rates with the calculations by one of the authors \cite{gal} that 
update the calculations by Motoba {\it et al.} \cite{motoba2,motoba3,motoba1}. 
These two sets of spectroscopic calculations agree reasonably well with each 
other for all hypernuclei considered in the present report, except 
for ${^{15}_{\Lambda}\mathrm{N}}$ which is discussed in detail below. 
The measured spectra are consistent with the 
observation, made in these shell-model calculations, that the partial decay 
contributions from the high-lying continuum of the daughter nuclear system 
outside the $0\hbar\omega$ $p$-shell configuration are unimportant 
in this mass range. The level of agreement between the reported measurements 
and the calculations allows us to confirm the previous spin-parity assignments 
made for ${^{7}_{\Lambda}\mathrm{Li}}$ and ${^{11}_{\Lambda}\mathrm{B}}$, 
and to assign $J^{\pi}={3/2}^+$ to ${^{15}_{\Lambda}\mathrm{N}}$ ground-state.

\section{Experimental and analysis techniques} 
\label{exper} 

FINUDA is a hypernuclear physics experiment, with cylindrical symmetry, 
installed at one of the two 
interaction regions of the $DA \Phi NE$ $e^{+}e^{-}$ collider, the INFN-LNF 
$\Phi$-factory. A description of the experimental apparatus can be found 
in \cite{fnd,fnd2}. Here we briefly sketch its main components, moving 
outwards from the beam axis: the {\it interaction/target region}, composed 
by a barrel of 12 thin scintillator slabs (TOFINO), surrounded by an octagonal 
array of $\mathrm{Si}$ microstrips (ISIM) facing eight target tiles; the 
{\it external tracking device}, consisting of four layers of position 
sensitive detectors (a decagonal array of $\mathrm{Si}$ microstrips (OSIM), 
two octagonal layers of low mass drift chambers (LMDC) and a stereo system of 
straw tubes (ST)) arranged in coaxial geometry; the {\it external time of 
flight detector} (TOFONE), a barrel of 72 scintillator slabs. The whole 
apparatus is placed inside a uniform 1.0 T solenoidal magnetic field; the 
tracking volume is immersed in $\mathrm{He}$ atmosphere to minimize the 
multiple scattering effect. 

The scientific program of the experiment is focussed on the study of 
spectroscopy and decay of $\Lambda$-hypernuclei produced by means of the 
$(K^{-}, \pi^{-})$ reaction with $K^{-}$'s at rest: 
\begin{equation} 
K^{-}_{stop} + {^{A}\mathrm{Z}} \rightarrow \pi^{-} + 
{^{A}_{\Lambda}\mathrm{Z}} 
\label{s-ex} 
\end{equation} 
by stopping in very thin targets the low energy ($\sim 16$ MeV) $K^{-}$'s 
coming from the $\Phi\rightarrow K^{-}K^{+}$ decay channel. In (\ref{s-ex}) 
${^{A}\mathrm{Z}}$ indicates the target nucleus and 
${^{A}_{\Lambda}\mathrm{Z}}$ the produced $\Lambda$-hypernucleus. 
$\Lambda$-hypernuclei decay through both the mesonic weak decay processes: 
\begin{eqnarray} 
^{A}_{\Lambda}\mathrm{Z} & \rightarrow & ^{A}(\mathrm{Z+1}) + \pi^{-} \\ 
^{A}_{\Lambda}\mathrm{Z} & \rightarrow & ^{A}\mathrm{Z} + \pi^{0} 
\label{mwd} 
\end{eqnarray} 
and the non-mesonic weak decay processes: 
\begin{eqnarray} 
^{A}_{\Lambda}\mathrm{Z} & \rightarrow & ^{A-2}(\mathrm{Z-1}) + p + n  \\ 
^{A}_{\Lambda}\mathrm{Z} & \rightarrow & ^{A-2}\mathrm{Z} + n + n  
\label{nmwd} 
\end{eqnarray}
where the final nuclear states in (6-9) are not necessarily particle stable. 
In contrast to the mesonic decays, the non-mesonic decays are not Pauli 
blocked, producing high-momentum nucleons ($\le 600$ MeV/c). 

The thinness of the target materials needed to stop the $K^{-}$'s, 
the high transparency of the FINUDA tracker and the very large solid angle 
($\sim 2\pi$ sr) covered by the detector ensemble make the FINUDA apparatus 
suitable to study the formation and the decay of $\Lambda$-hypernuclei 
by means of high resolution magnetic spectroscopy of the charged particles 
emitted in the processes (\ref{s-ex}) \cite{fnd}, (6) and (8) \cite{npa804}; 
the features of the apparatus give also the possibility to investigate 
many other final states produced in the 
interaction of stopped kaons with nuclei \cite{kpp}. 

In this paper results are presented obtained by analyzing the data collected 
by the FINUDA experiment from 2003 to 2007 with a total integrated luminosity 
of $1156\ pb^{-1}$. Only targets leading to the formation of the $p$-shell 
hypernuclei $^{7}_{\Lambda}\mathrm{Li}$, $^{9}_{\Lambda}\mathrm{Be}$, 
$^{11}_{\Lambda}\mathrm{B}$ and $^{15}_{\Lambda}\mathrm{N}$ are here 
considered, namely $^{7}\mathrm{Li}$ (2$\times$, $4$ mm thick, natural 
isotopic composition), $^{9}\mathrm{Be}$ (2$\times$, $2$ mm thick, natural 
isotopic composition), $^{12}\mathrm{C}$ (3$\times$, $1.7$ mm thick, natural 
isotopic composition, mean density $2.265\ g\ cm^{-3}$) and 
$\mathrm{D}_{2}\mathrm{O}$ (mylar walled, 1$\times$, $3$ mm thick), together 
with $^{6}\mathrm{Li}$ targets (2$\times$, $4$ mm thick, $90\%$ enriched) 
leading to the production of $^{5}_{\Lambda}\mathrm{He}$; 
$^{5}_{\Lambda}\mathrm{He}$ is reported for the sake of completeness. 

To investigate the  MWD process (6) events where analyzed in which two 
$\pi^{-}$'s were detected in coincidence. One $\pi^{-}$, with a momentum 
as high as $260-290$ MeV/c, gives the signature of the formation of the 
ground-state of the hypernuclear system or of a low lying excited state 
decaying to it by electromagnetic emission. The second $\pi^{-}$, with 
a momentum lower than 115 MeV/c, gives the signature of the decay. 
By requiring this coincidence, negative pions are the only negative particles 
originating from the $K^{-}$'s stopping point in the targets that enter the 
tracking volume of the apparatus. Nevertheless, to get a cleaner data sample, 
only tracks identified as $\pi^{-}$'s by the FINUDA detectors were considered. 
In particular, the information of the specific energy loss in both OSIM 
and the LMDC's and the mass identification from the time of flight system 
(TOFINO-TOFONE), if present, were used to obtain a multiple identification 
selection. 

In the present analysis we required good quality tracks to determine 
the momentum of the formation $\pi^{-}$. These tracks must originate 
in a properly defined fiducial volume around the primary $K^{-}$ vertex 
and are identified by four hits, one in each of the FINUDA tracking detectors 
({\it long tracks}), and are selected with a quite strict requirement on 
the $\chi^{2}$ from the track fitting procedure (corresponding to a $90\%$ 
confidence level). They have a resolution $\Delta p /p\sim 1\%$ FWHM in the 
region 260$-$280 MeV/c; this resolution is about twice worse than the best 
value obtained with top quality tracks \cite{fnd} for spectroscopy studies. 
The worsening was due to the more relaxed quality criteria applied to 
increase the statistics of the sample available for the coincidence 
measurement. In particular, no cut has been made to select the direction 
of the outgoing tracks. 

\begin{table}[h] 
\begin{center} 
\begin{tabular}{|c|c|c|c|c|} 
\hline 
target & final & $p_{\pi^{-}}$ & B.E. ($_{\Lambda}^{A}{\rm Z}$)&references \\ 
$^{A}{\rm Z}$  & hypernucleus & (MeV/c) & (MeV) & \\  \hline 
$^{6}\mathrm{Li}$ & $^{5}_{\Lambda}\mathrm{He}$ & 272 $-$ 278 & 0.63 $-$ 5.99 
&  \cite{szym,kame} \\ \hline 
$^{7}\mathrm{Li}$ & $^{7}_{\Lambda}\mathrm{Li}$ & 273 $-$ 279 & 1.85 $-$ 7.45 
& \cite{juric, hashim, hyp06} \\  \hline 
$^{9}\mathrm{Be}$ & $^{9}_{\Lambda}\mathrm{Be}$ & 280 $-$ 286 & 1.50 $-$ 7.00 
& \cite{juric, hashim, noumi2,pile} \\  \hline 
$^{12}\mathrm{C}$ & $^{11}_{\Lambda}\mathrm{B}$ & 258 $-$ 264 & $-$2.00 $-$ 
2.75 & \cite{pile,hotchi, fnd} \\  \hline 
$^{12}\mathrm{C}$ & $^{12}_{\Lambda}\mathrm{C}$ & 267 $-$ 273 & 9.00 $-$ 14.00 
& \cite{pile,hotchi,fnd} \\  \hline 
$^{16}\mathrm{O}$ & $^{15}_{\Lambda}\mathrm{N}$ & 265 $-$ 270.5 & 0.0 $-$ 
4.90 & \cite{pile,hashim} \\  \hline 
$^{16}\mathrm{O}$ & $^{16}_{\Lambda}\mathrm{O}$ & 270.5 $-$ 282 
& 4.90 $-$ 15.40 & \cite{pile,hashim} \\  \hline 
\end{tabular} 
\caption{Summary of the momentum and binding energy (B.E.) intervals selected 
to identify the formation of various hypernuclear systems. First column: 
target nucleus $^{A}{\rm Z}$; second column: weakly decaying final 
hypernucleus; third column: production-pion momentum interval; fourth column: 
$_{\Lambda}^{A}{\rm Z}$ binding energy interval; fifth column: references to 
previous missing mass spectroscopy experiments.} 
\label{tab1} 
\end{center} 
\end{table} 

Table~\ref{tab1} reports the binding energy intervals selected to identify 
the formation of the different hypernuclei. 
The intervals have been determined by comparing our experimental inclusive 
formation spectra with the known values of binding energies for ground-states 
and low lying excited states, as deduced from the references indicated in 
the last column. The interval width takes into account our experimental 
resolution, $\sigma_{\mathrm{p}} \sim 1$ MeV/c and 
$\sigma_{\mathrm{B.E.}} \sim 1$ MeV for a typical pion momentum of about 
270 MeV/c. For $^{7}_{\Lambda}\mathrm{Li}$ a sharp cut was set at an 
excitation energy of 3.94 MeV, corresponding to the threshold for the 
${^{7}_{\Lambda}\mathrm{Li}} \rightarrow {^{5}_{\Lambda}\mathrm{He}} + d$ 
fragmentation. As it is well known, for $^{11}_{\Lambda}\mathrm{B}$ and 
$^{15}_{\Lambda}\mathrm{N}$, produced on $^{12}\mathrm{C}$ and 
$^{16}\mathrm{O}$ ($\mathrm{D}_{2}\mathrm{O}$) targets respectively, 
the production momentum region {\it partially} overlaps the higher part of 
the momentum spectrum of $\pi^{-}$'s emitted in the $\Lambda$ quasi-free 
($\Lambda_{qf}$) production. This holds particularly for 
$^{15}_{\Lambda}\mathrm{N}$ which is expected to be dominantly formed by 
proton emission from the two peaks of $^{16}_{\Lambda}\mathrm{O}$ observed 
at B.E.$\simeq 2$ MeV and B.E.$\simeq -4$ MeV \cite{boh3}. However, in order 
to minimize the contamination in the $^{15}_{\Lambda}\mathrm{N}$ decay 
spectrum by decays of other hypernuclear species, which may be formed in the 
opening of higher energy emission channels, events were selected corresponding 
only to the positive B.E. $^{16}_{\Lambda}\mathrm{O}$ peak, as indicated in 
Table~\ref{tab1}. A likely source of contamination is provided by the 
production of $^{12}_{\Lambda}\mathrm{C}$ and its subsequent mesonic decay: 
\begin{eqnarray} 
K^- + {^{16}\mathrm{O}} & \rightarrow & \pi^- + \alpha + 
{^{12}_{\Lambda}\mathrm{C}}  \nonumber \\ 
^{12}_{\Lambda}\mathrm{C}&\rightarrow&^{12}\mathrm{N}+\pi^{-}\ 
(p_{\pi^{-}~max}\simeq 91\ \rm{MeV/c}). 
\label{frag} 
\end{eqnarray} 
For $^{11}_{\Lambda}\mathrm{B}$, on the other hand, it was enough to focus on 
the sizable excitation peak of $^{12}_{\Lambda}\mathrm{C}$ at B.E.$\simeq 0$ 
MeV which is known to lead, upon proton emission, to several 
excited states of $^{11}_{\Lambda}\mathrm{B}$ \cite{dalitzB11}. 

\begin{figure}[h] 
\begin{center} 
\includegraphics[width=0.5\textwidth,height=5cm]{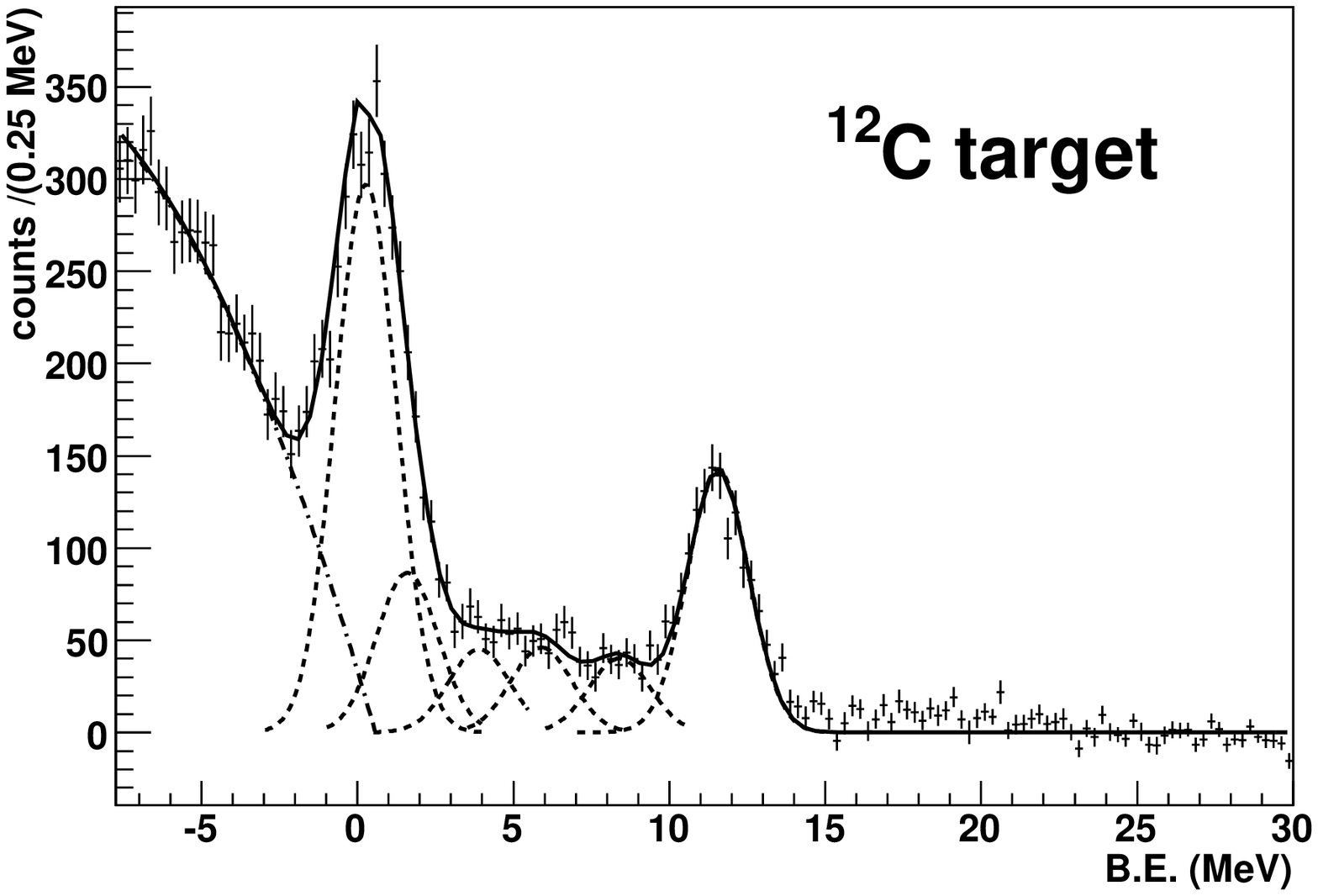} 
\hspace{-2.5mm} 
\includegraphics[width=0.5\textwidth,height=5.1cm]{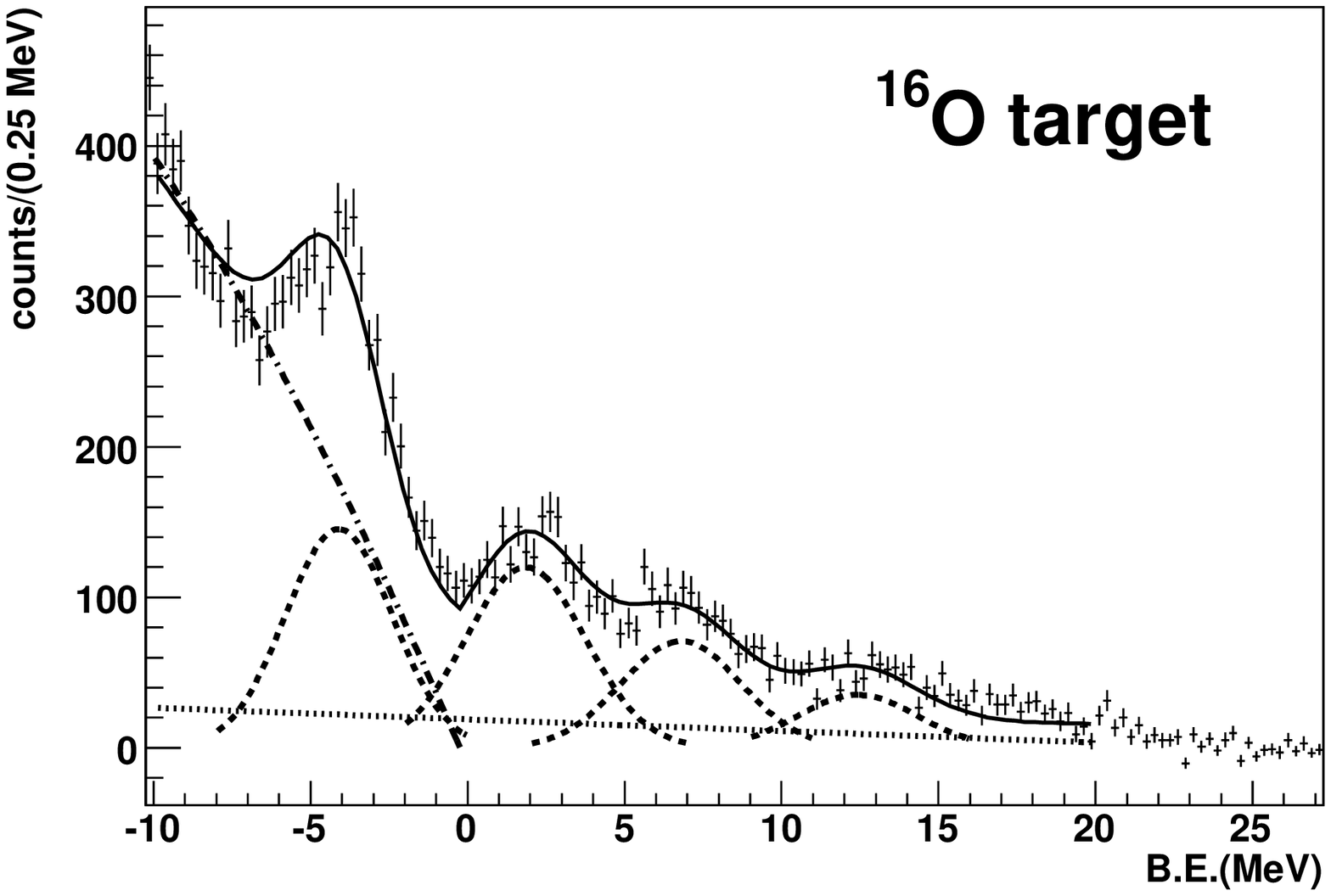} 
\caption{Inclusive binding energy spectra for good quality $\pi^-$ tracks 
coming from $^{12}\mathrm{C}$ (left) and $^{16}\mathrm{O}$ (right) targets. 
The continuous line is the best fit curve to the spectra; the dashed curves 
represent the contributions from the known hypernuclear states and the 
dot-dashed curve the $\Lambda_{qf}$ background. For $^{16}\mathrm{O}$ the 
dotted curve parametrizes the $K^{-}$ decay in flight.} 
\label{fig0} 
\end{center} 
\end{figure} 

Moreover, it should be noted that the contribution to the inclusive spectra 
due to the reaction chain: 
\begin{equation} 
K^{-} + (np) \rightarrow \Sigma^{-} + p \ \ \ \ \ \ \ \ \Sigma^{-} 
\rightarrow n + \pi^{-} 
\label{Knp} 
\end{equation} 
constitutes the only physical background below the hypernuclear formation 
peaks. It was evaluated by simulating a sample of background events and 
applying to the simulated data the same selection criteria as for the real 
ones. The background spectra were then normalized to the experimental ones 
above the kinematical limits for the hypernuclear ground-state formation 
and subtracted. A detailed description of such a procedure is available in 
Ref.~\cite{npa804}. 

Figure \ref{fig0} shows the inclusive binding energy spectra for formation 
$\pi^-$ from $^{12}\mathrm{C}$ and $^{16}\mathrm{O}$ targets, after 
subtraction of the $K^{-}(np)$ background. The continuous line is the best 
fit to the spectra, while the dashed curves represent the contributions 
from the known hypernuclear states and the dot-dashed curve represents the 
polynomial background, due to $\Lambda_{qf}$ production in the negative B.E. 
region. In the positive B.E. region a background contribution from $K^{-}$'s 
decay in flight, shown separately in the figure for $^{16}\mathrm{O}$, is 
also considered. This background affects differently the targets placed in 
different positions with respect to the $e^{+}e^{-}$ interaction region due 
to the fact that the $(e^{+},e^{-})$ crossing beams collide with an angle of 
12.5 mrad in order to increase the luminosity; the effect has been very well 
studied in \cite{fnd2}. On the other hand, this background affects only the 
inclusive B.E. spectra and does not give any contribution to the low energy 
$\pi^{-}$ spectra from MWD, for which a two $\pi^{-}$ coincidence is required. 

Table~\ref{tab2} reports the binding energy values of the hypernuclear 
states obtained from the global fitting procedure in the B.E. regions 
indicated in Table~\ref{tab1}, as mean values of the corresponding 
gaussians; the $\chi^2/ndf$ values are also indicated. It must be noted 
that these binding energy values can be different from the ones obtained 
in analyses dedicated to spectroscopy studies, due to the relaxation of 
the quality cuts applied to the tracks: indeed, the requirement of the 
coincidence with a low momentum $\pi^{-}$ acts as a filter that allows to 
untighten the quality selections on the long tracks, as will be shown in 
the next section. The precise choice of peaks does not change the MWD spectra 
at any qualitative level, except for reducing the sample statistics. 

\begin{table}[h] 
\begin{center} 
\begin{tabular}{|c|c|c|c|c|c|} 
\hline 
peak &  &$^{7}_{\Lambda}\mathrm{Li}$  & $^{9}_{\Lambda}\mathrm{Be}$ & 
$^{12}_{\Lambda}\mathrm{C}$ &  $^{16}_{\Lambda}\mathrm{O}$ \\ 
     & &  (MeV) & (MeV)  & (MeV)  & (MeV) \\  \hline 
1 g.s.& B.E. & $5.85 \pm 0.13$ & $6.30 \pm 0.10$ & $11.57 \pm 0.04$ & 
12.42 fixed \\ \hline 
2  &  B.E & $3.84 \pm 0.15$ & $3.45 \pm 0.10$ & 8.4 fixed & 
$6.800 \pm 0.017$ \\ \hline 
3 &B.E.& $1.9 \pm 0.3$ & $0.25 \pm 0.22$ & 5.9 fixed & $1.85$ fixed \\ \hline 
4 & B.E & $0.39\pm 0.20$ & & 3.9 fixed & $-4.100\pm 0.004$ \\ \hline 
5 & B.E  & $-2.000 \pm 0.047$ &  & 1.6 fixed & \\ \hline 
6  & B.E   &           &   & 0.27 fixed & \\ \hline 
& $\chi^2/ndf$ & 1.10  & 1.00  & 1.72  &  1.78\\  \hline 
\end{tabular} 
\caption{Mean values of the gaussians representing hypernuclear 
states in global best fits to binding energy inclusive spectra for 
$^{7}_{\Lambda}\mathrm{Li}$, $^{9}_{\Lambda}\mathrm{Be}$, 
$^{12}_{\Lambda}\mathrm{C}$ and $^{16}_{\Lambda}\mathrm{O}$: 
only the peaks contributing to the B.E. selections of Table~\ref{tab1} 
are considered. The FWHM of the peaks for $^{7}_{\Lambda}\mathrm{Li}$, 
$^{9}_{\Lambda}\mathrm{Be}$ and $^{12}_{\Lambda}\mathrm{C}$ is $2.31$ MeV, 
while for $^{16}_{\Lambda}\mathrm{O}$ is $4.48$ MeV, due to the 
malfunctioning of the outer drift chamber directly facing the target. 
Values of $\chi^2/ndf$ for global fits (hypernuclear states and polynomial 
background) are also reported. See the references in the fifth column of 
Table~\ref{tab1} for comparison with previous measurements.} 
\label{tab2} 
\end{center} 
\end{table} 

For the decay $\pi^{-}$ momentum measurement only tracks not reaching the 
ST system ({\it short tracks}) have been used. The lower threshold for the 
detection momentum of these $\pi^{-}$'s is $\sim 80$ MeV/c. These tracks 
correspond mainly to particles 
{\it backward emitted} from the targets, crossing the whole 
interaction/target region before entering the tracker; 
their momentum resolution is $\Delta p / p \sim 6\%$ FWHM at 110 MeV/c. 

The acceptance for low energy $\pi^{-}$'s, $\epsilon$, was evaluated for each 
target, taking into account the geometrical layout, the efficiency of the 
FINUDA pattern recognition algorithm, the trigger and the efficiency of 
the quality cuts applied in the analysis procedure. The acceptance function, 
$R = 1/\epsilon$, for the momentum features a negative quadratic exponential 
behaviour in the $80-160$ MeV/c range and flattens above $90$ MeV/c; for the 
kinetic energy the behaviour is similar in the $20-70$ MeV range, as shown 
in Fig.~\ref{fig1} for $^{7}_{\Lambda}\mathrm{Li}$ with a dot-dashed 
line. The error on the acceptance function is always $< 5\%$.

\section{MWD $\pi^{-}$'s spectra} 
\label{p_mwd} 

It is worth recalling that the information available up to now on 
the charged MWD of light hypernuclei consists almost entirely of 
$\Gamma_{\pi^{-}} / \Gamma_{\Lambda}$ values obtained by means of counting 
measurements in coincidence with the hypernuclear formation $\pi^{-}$ 
detection, with no magnetic analysis of the decay meson; $\pi^{-}$ kinetic 
energy spectra have been obtained for $^{12}_{\Lambda}\mathrm{C}$ MWD 
only \cite{sato}. The $\pi^{-}$ spectra presented here allow to have a more 
careful confirmation of the elementary mechanism that is supposed to underlie 
the decay process, as well as to have information on the spin-parity of the 
initial hypernuclear ground state. In this respect the study of pion spectra 
from MWD can be regarded as an indirect spectroscopic investigation tool. 

Due to the $\pi^{-}$ momentum detection threshold of the apparatus 
($\sim\ 80$ MeV/c), only MWD spectra of $^{7}_{\Lambda}\mathrm{Li}$, 
$^{9}_{\Lambda}\mathrm{Be}$, $^{11}_{\Lambda}\mathrm{B}$ ($^{12}\mathrm{C}$ 
targets) and $^{15}_{\Lambda}\mathrm{N}$ ($^{16}\mathrm{O}$ target) were 
investigated. Spectra from $^{12}_{\Lambda}\mathrm{C}$ and 
$^{16}_{\Lambda}\mathrm{O}$ could not be observed. 

Background coming from $\Lambda_{qf}$ decay was simulated taking into account 
the Fermi momentum of the neutron in the target: it was found that the spectrum 
of the decay $\pi^-$ momentum extends up to $\sim$ 160 MeV/c, well above the 
stopping point of the hypernucleus MWD  contribution at $\sim$ 110 MeV/c. 
This background was then subtracted from the $^{11}_{\Lambda}\mathrm{B}$ spectrum 
by normalizing the area of the simulated spectra, after reconstruction, 
to the experimental ones in the $110-160$ MeV/c decay-pion region, 
populated only by $\Lambda_{qf}$ decays. 
Each spectrum was corrected by means of the acceptance function, $R$, 
described in the previous section. The decay $\pi^{-}$ momentum spectra 
show interesting structures whose meaning can be better understood 
by considering the corresponding kinetic energy spectra that are directly 
related to the excitation function of the daugther nucleus. Kinetic energy 
spectra, background subtracted and acceptance corrected, were evaluated 
for MWD of $^{7}_{\Lambda}\mathrm{Li}$, $^{9}_{\Lambda}\mathrm{Be}$, 
$^{11}_{\Lambda}\mathrm{B}$ and $^{15}_{\Lambda}\mathrm{N}$. 

\begin{figure}[h] 
\begin{center} 
\includegraphics[width=95mm]{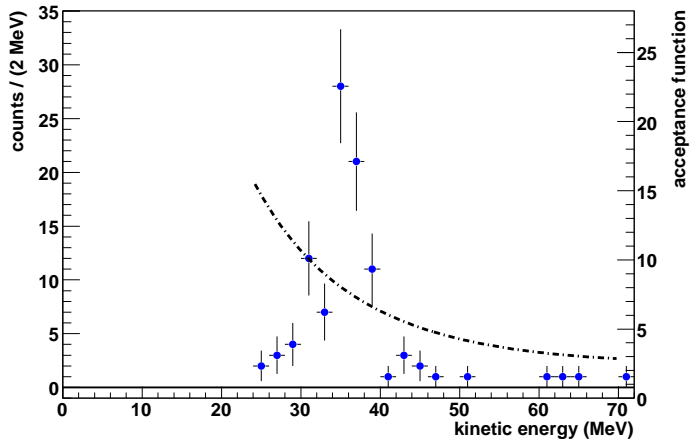} 
\caption{Kinetic energy spectrum of MWD $\pi^{-}$ for 
$^{7}_{\Lambda}\mathrm{Li}$ before acceptance correction. The dot-dashed 
curve is the acceptance function, $R$, to be applied to the data.} 
\label{fig1} 
\end{center} 
\end{figure} 

In Fig. \ref{fig1}, the kinetic energy spectrum for 
$^{7}_{\Lambda}\mathrm{Li}$, 
before applying the acceptance correction, is reported. 
The acceptance correction is represented by the dot-dashed curve. 
The errors are statistical only. It is evident that the 
low energy $\pi^{-}$ spectrum is practically background free and that the 
$50-70$ MeV contribution is negligible and compatible with zero. 
This demonstrates the effectiveness of the coincidence requirement. 
The small residual background, which is similar also for the other targets, 
has been evaluated after acceptance correction and subtracted to calculate 
the decay ratio (see next section). In the following only the $16-60$ MeV 
region of the MWD spectra will be shown.

\subsection{$^{7}_{\Lambda}\mathrm{Li}$} 

\begin{figure}[h] 
\begin{center} 
\includegraphics[width=90mm]{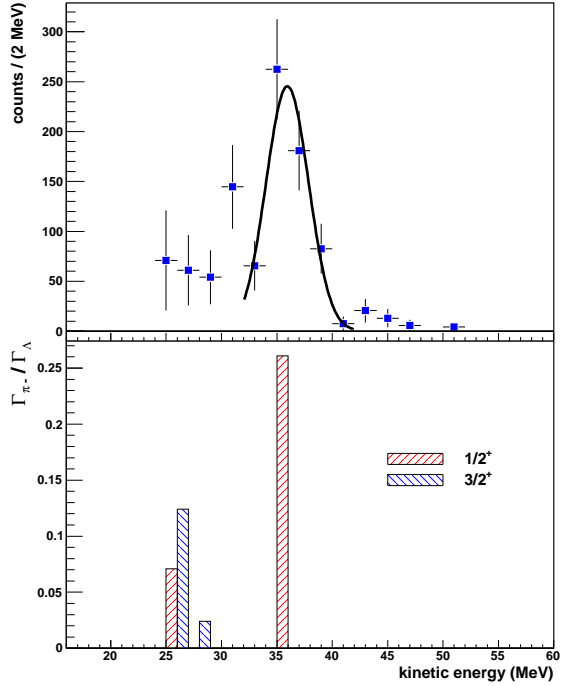} 
\vspace{-5mm} 
\caption{Upper part: kinetic energy spectrum of MWD $\pi^{-}$ for 
$^{7}_{\Lambda}\mathrm{Li}$ after acceptance correction. The solid curve 
is a gaussian fit to the peak in the spectrum, to compare with theoretical 
predictions in the lower part. Lower part: calculated major decay rates 
to final $^{7}\mathrm{Be}$ states from \cite{gal}, in red bars for 
$^{7}_{\Lambda}\mathrm{Li}$ ground-state spin-parity $1/2^{+}$, and in blue 
bars for $^{7}_{\Lambda}\mathrm{Li}$ ground-state spin-parity $3/2^{+}$ 
(see text)}. 
\label{fig1_1} 
\end{center} 
\end{figure} 

In Fig. \ref{fig1_1} the acceptance corrected spectrum for 
$^{7}_{\Lambda}\mathrm{Li}$ is shown in the 
upper part and compared with calculated decay ratios 
($\Gamma_{\pi^{-}}/\Gamma_{\Lambda}$) to final $^{7}\mathrm{Be}$ states 
\cite{gal} shown in the lower part. These calculated 
rates are close to those calculated by Motoba {\it et al.} \cite{motoba1}. 
The errors in the spectrum of the upper part are inclusive of both the 
statistical and the acceptance contributions. Only major contributions 
are shown in the lower part with a common, arbitrary $1$ MeV width, although 
each of the two bars in Fig.~\ref{fig1_1} between 25 and 27 MeV stands for 
$^{7}\mathrm{Be}$ states spread over roughly $2-3$ MeV. The correspondence of 
the structures observed in the experimental spectra with the rates of decay 
to different excited states of the daugther nucleus, assuming initial 
spin-parity $1/2^{+}$, is clear. The peak 
structure corresponds to the production of $^{7}\mathrm{Be}$ in its $3/2^{-}$ 
ground-state and in its only bound $1/2^{-}$ excited state, at $429$ keV. 
Due to the FINUDA experimental resolution these close levels are not resolved 
and the gaussian fit superimposed on the data points yields a FWHM $\sim$ 4.5 
MeV, compatible with the intrinsic resolution of the apparatus: 
$\Delta T / T \sim 11\%$ FWHM at 38 MeV. The part of the spectrum at lower 
energies is due to three body decays. 

The portion of the spectrum that cannot be measured due to our experimental 
detection threshold ($\sim 22$ MeV) is negligible if compared with the errors 
affecting the counts in the spectrum, according to the calculated excitation 
spectra shown in \cite{motoba1}. The same holds also for the other hypernuclei 
studied in this paper. It is then reasonable to compare the total area of the 
spectrum with the decay rates summed over the whole excitation 
energy interval, as done in the next section. 

The shape of the spectrum confirms the spin assigned to the hypernuclear 
ground state of $^{7}_{\Lambda}\mathrm{Li}$ \cite{sasao}. Indeed, only 
a spin-parity $1/2^+$ for $^{7}_{\Lambda}\mathrm{Li}$ ground state, shown by 
red bars, reproduces the fitted peak at $\sim 36$ MeV due to the 
$^{7}\mathrm{Be}$ ground state and excited state at 429 keV. A spin-parity 
$3/2^+$ for $^{7}_{\Lambda}\mathrm{Li}$ ground state would imply a radically 
different spectrum shape \cite{motoba1,gal}, as indicated in Fig.~\ref{fig1_1} 
by the blue bars, short of any peak about  $^{7}\mathrm{Be}$ ground state and 
its 429 keV excited state.

\subsection{$^{9}_{\Lambda}\mathrm{Be}$} 
\begin{figure}[h] 
\vspace{-8mm} 
\begin{center} 
\includegraphics[width=90mm]{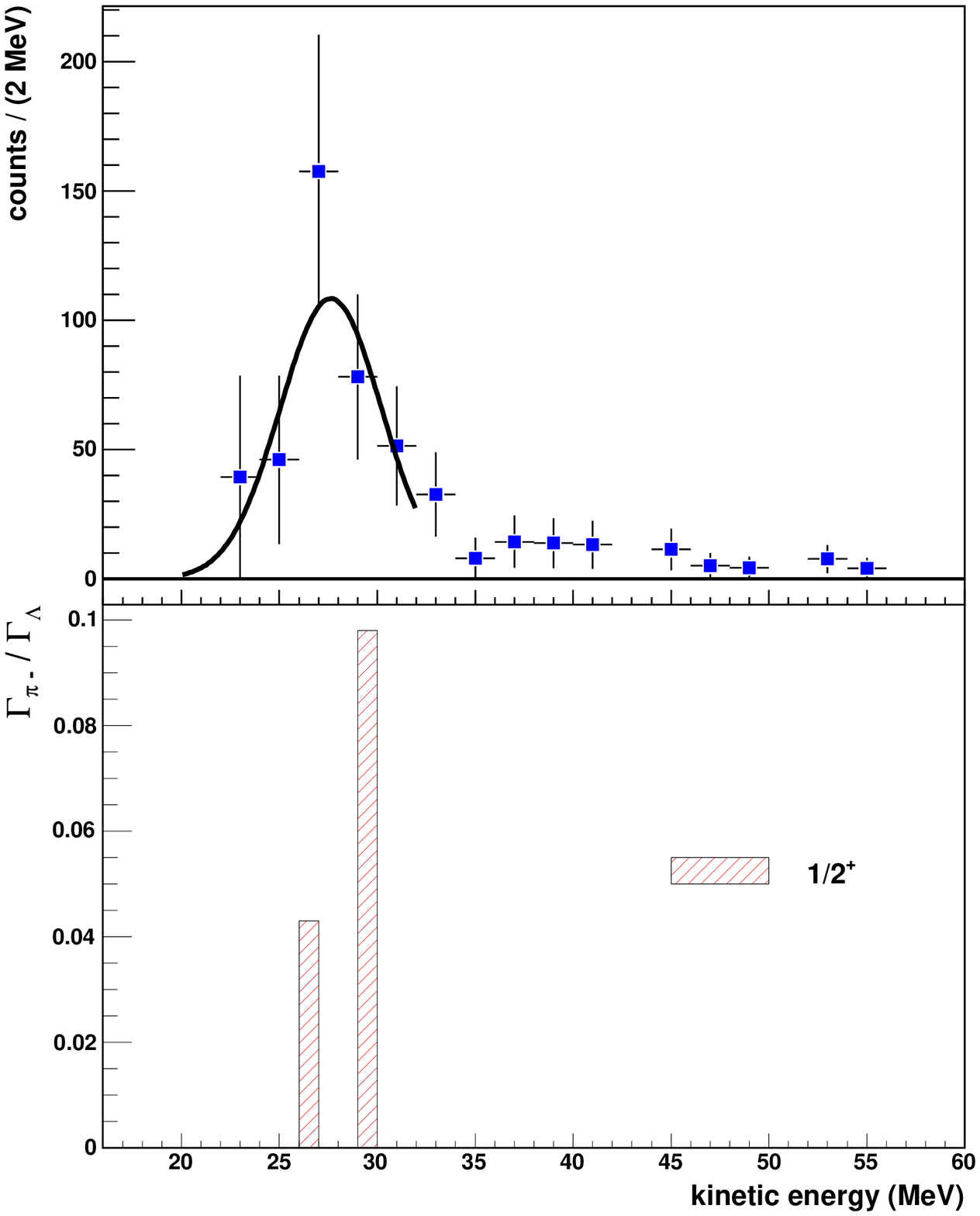} 
\vspace{-5mm} 
\caption{Upper part: kinetic energy spectrum of MWD $\pi^{-}$ for 
$^{9}_{\Lambda}\mathrm{Be}$ after acceptance correction. The solid curve 
is a gaussian fit to the peak in the spectrum, to compare with theoretical 
predictions in the lower part. Lower part: calculated major decay rates to 
final $^{9}\mathrm{B}$ states from \cite{gal}, in red bars for 
$^{9}_{\Lambda}\mathrm{Be}$ ground-state spin-parity $1/2^{+}$.} 
\label{fig1a} 
\end{center} 
\end{figure} 

In Fig.~\ref{fig1a} the acceptance corrected spectrum for 
$^{9}_{\Lambda}\mathrm{Be}$ is shown in the upper part and compared with 
calculated decay ratios ($\Gamma_{\pi^{-}}/\Gamma_{\Lambda}$) to final 
$^{9}\mathrm{B}$ states \cite{gal} shown in the lower part. The errors in 
the spectrum of the upper part are again inclusive of both the statistical 
and the acceptance contributions. Only major contributions are shown in the 
lower part with a common, arbitrary $1$ MeV width. In this case too, as for 
$^{7}_{\Lambda}\mathrm{Li}$ above, the calculated rates are very close to 
those calculated by Motoba {\it et al.} \cite{motoba1}. 

In the $^{9}_{\Lambda}\mathrm{Be}$ spectrum our energy resolution does not 
allow a separation between the two components predicted to dominate the 
spectrum \cite{motoba1,gal}, the $^{9}\mathrm{B}$ ground-state $3/2^{-}$ and 
the excited state $1/2^{-}$ at $2.75$ MeV. As a consequence, the gaussian fit 
superimposed on the data points yields a FWHM $\sim$ 7.5 MeV. 
The correspondence between the experimental spectrum and the calculated rates 
of decay to different excited states of the daugther nucleus is clear. 
Our spectrum is consistent with the interpretation from $(\pi^+,K^+)$ reactions 
\cite{hashim} according to which the $^{9}_{\Lambda}\mathrm{Be}$ ground state 
is dominantly a $1s$-$\Lambda$ coupled to $^{8}\mathrm{Be}(0^{+})$ ground 
state.

\subsection{$^{11}_{\Lambda}\mathrm{B}$} 
\begin{figure}[h] 
\vspace{-8mm} 
\begin{center} 
\includegraphics[width=90mm]{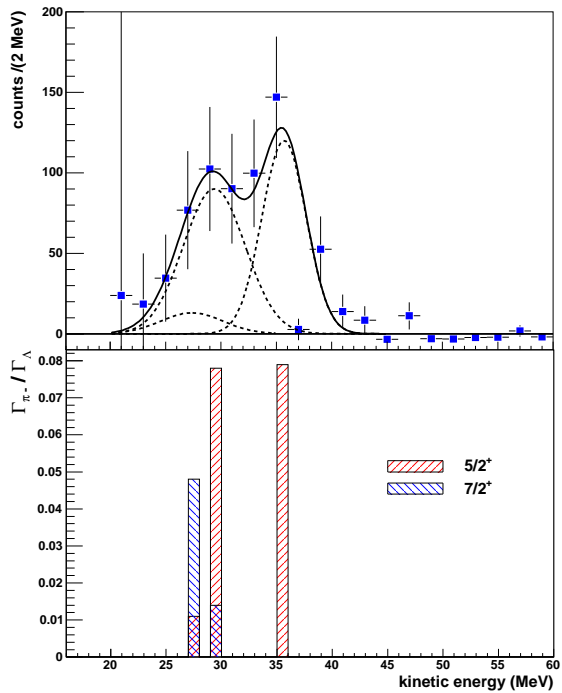} 
\vspace{-5mm} 
\caption{Upper part: kinetic energy spectrum of MWD $\pi^{-}$ for 
$^{11}_{\Lambda}\mathrm{B}$ after acceptance correction. The solid curve is 
a two-gaussian fit to the peaks in the spectrum, to compare with theoretical 
predictions in the lower part; dashed curves are the single components. Lower 
part: calculated major decay rates to final $^{11}\mathrm{C}$ states from 
\cite{gal}, in red bars for $^{11}_{\Lambda}\mathrm{B}$ ground-state 
spin-parity $5/2^{+}$, and in blue bars for $^{11}_{\Lambda}\mathrm{B}$ 
ground-state spin-parity $7/2^{+}$.} 
\label{fig2} 
\end{center} 
\end{figure} 

In Fig. \ref{fig2} the spectrum for $^{11}_{\Lambda}\mathrm{B}$ is shown 
and compared with calculated decay rates to final $^{11}\mathrm{C}$ 
states \cite{gal}. The errors in the spectra are again inclusive of both the 
statistical and the acceptance contributions. Major and secondary 
contributions are shown in the lower part with a common, arbitrary $1$ MeV 
width. Assuming ground-state spin-parity $5/2^{+}$, it is possible to identify 
two major contributions in the $^{11}_{\Lambda}\mathrm{B}$ spectrum due to 
$^{11}\mathrm{C}$ ground-state $3/2^{-}$ and its $7/2^{-}$ excited state at 
6.478 MeV, both shown by red bars. A third contribution due to the $3/2^{-}$ 
excited state at 8.10 MeV is considerably weaker than the former ones, 
its main effect being to introduce some asymmetry to the overall spectrum 
towards lower kinetic energies. Additional contribution of a similar 
strength \cite{motoba2} in this energy range arises from transitions to 
several $sd$ states within $7-10$ MeV excitation energy in $^{11}\mathrm{C}$ 
(not shown in Fig.~\ref{fig2}). It is clear from the figure that the shape 
of the spectrum is well reproduced by assigning spin-parity $5/2^{+}$ to 
$^{11}_{\Lambda}\mathrm{B}_{g.s.}$. 
Assuming ground-state spin-parity $7/2^{+}$, the $^{11}\mathrm{C}$ 
ground-state peak is missing and the dominant decay is to the $5/2^{-}$ 
excited state at 8.420 MeV shown by a blue bar. A secondary contribution due to 
the $7/2^{-}$ excited state at 6.478 MeV is also considerably weaker than that 
arising under the assumption of spin-parity $5/2^{+}$. We note that the major 
contributions to the $^{11}_{\Lambda}\mathrm{B}$ spectrum discussed above for 
both $^{11}_{\Lambda}\mathrm{B}_{g.s.}$ possible spin-parity values are also 
borne out by the calculation of Ref.~\cite{motoba2}. 

A $5/2^{+}$ assignment for $^{11}_{\Lambda}\mathrm{B}$ ground-state, 
first made by Zieminska studying emulsion spectra \cite{ziem}, was 
experimentally confirmed by the KEK measurement \cite{sato} comparing 
the derived value of the total $\pi^-$ decay rate with the total $\pi^-$ 
decay rate calculated in \cite{motoba2}. 
The present measurement of the decay spectrum shape provides a 
confirmation of $J^{\pi}(^{11}_{\Lambda}\mathrm{B}_{g.s.}) = 5/2^{+}$ 
by a different observable.

\subsection{$^{15}_{\Lambda}\mathrm{N}$} 
\begin{figure}[h] 
\vspace{-10mm} 
\begin{center} 
\includegraphics[width=90mm]{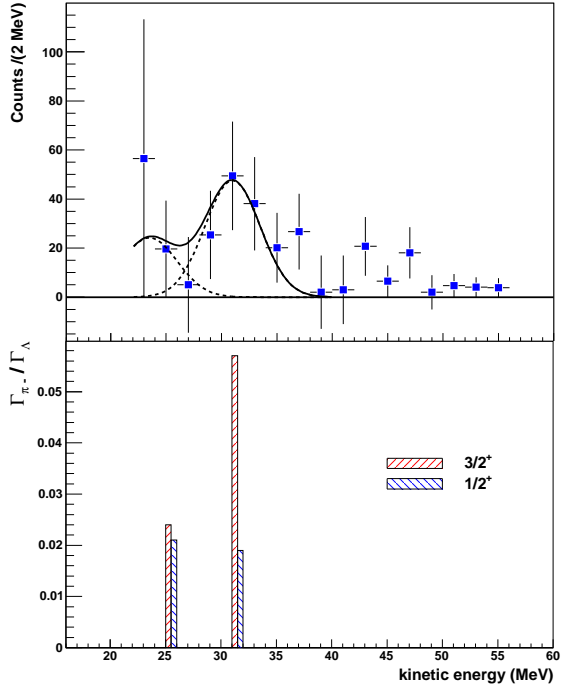} 
\vspace{-5mm} 
\caption{Upper part: kinetic energy spectrum of MWD $\pi^{-}$ for 
$^{15}_{\Lambda}\mathrm{N}$ after acceptance correction. The solid curve is 
a two-gaussian fit to the peaks in the spectrum, to compare with theoretical 
predictions in the lower part; dashed curves are the single components. 
Lower part: calculated major decay rates to final $^{15}\mathrm{O}$ states 
from \cite{gal}, in red bars for $^{15}_{\Lambda}\mathrm{N}$ ground-state 
spin-parity $3/2^{+}$, and in blue bars for $^{15}_{\Lambda}\mathrm{N}$ 
ground-state spin-parity $1/2^{+}$. Note that in this case the arbitrary bar 
width was reduced to 0.5 MeV to avoid overlap and to facilitate the comparison 
between the two spin hypotheses: indeed, the energies of the produced final 
states are practically the same.} 
\label{fig2a} 
\end{center} 
\end{figure} 

In Fig. \ref{fig2a} the spectrum for $^{15}_{\Lambda}\mathrm{N}$ is shown 
and compared with calculated decay rates to final $^{15}\mathrm{O}$ 
states \cite{gal}. The errors in the spectra are again inclusive of both the 
statistical and the acceptance contributions. 
The contribution from $\Lambda_{qf}$ was evaluated to be less than $5\%$ and then 
neglected, taking into account the overwhelming importance of the statistical 
errors. 
In the experimental spectrum, 
the $^{15}\mathrm{O}$ ground-state $1/2^{-}$ contribution stands out clearly, 
along with a hint for a secondary structure separated by about 6 MeV. 
The gaussian fit 
of the ground state component yields a FWHM of $\sim$ 6 MeV, larger than our 
standard value $\sim$ 4.5 MeV due to the already mentioned malfunctioning of 
the apparatus (see caption of Table \ref{tab2}) and to the limited statistics. 
The fit to the lower-energy secondary structure is strongly influenced by the 
substantial error affecting the lowest energy point. 
According to Refs.~\cite{motoba3,gal}, this secondary structure derives most 
of its strength from $sd$ states scattered around 6 MeV excitation while the 
contribution of the $^{15}\mathrm{O}$ $p_{3/2}^{-1}p_{1/2}$ excited state at 
6.176 MeV is negligible. We note that, in the upper part of Fig.~\ref{fig2a}, 
the channel at about 37 MeV kinetic energy might get contribution from the reaction 
chain 
$K^{-} + {^{16}\mathrm{O}} \rightarrow \pi^{0} + \alpha + 
{^{12}_{\Lambda}\mathrm{B}}$, $^{12}_{\Lambda}\mathrm{B} \rightarrow \pi^{-} + 
{^{12}\mathrm{C}_{g.s.}}$, with a non negligible pion charge exchange in 
the $\mathrm{D}_{2}\mathrm{O}$ target.  

The ground-state spin has not been determined experimentally. The most recent 
theoretical study of $\Lambda$-hypernuclear spin dependence \cite{boh3} 
predicts $J^{\pi}(^{15}_{\Lambda}\mathrm{N}_{g.s.}) = 3/2^{+}$, setting the 
$1/2^{+}$ excited ground-state doublet level about 90 keV above the $3/2^{+}$ 
ground-state. The spin ordering, however, cannot be determined from the 
$\gamma$-ray de-excitation spectra measured recently on a $^{16}\mathrm{O}$ 
target at Brookhaven \cite{boh4}. As for MWD, the prominence of 
$^{15}\mathrm{O}_{g.s.}$ in the spectrum of Fig.~\ref{fig2a} supports this 
$J^{\pi}(^{15}_{\Lambda}\mathrm{N}_{g.s.}) = 3/2^{+}$ theoretical assignment 
since the decay $^{15}_{\Lambda}\mathrm{N}(1/2^{+}) \to \pi^- + 
{^{15}\mathrm{O}_{g.s.}}$ is suppressed according to the following simple 
argument. Recent shell-model calculations suggest that the nuclear-core 
$^{14}\mathrm{N}_{g.s.}$ wavefunction is very close to a $^{3}D_1$ 
wavefunction \cite{boh3}, which for $J^{\pi}(^{15}_{\Lambda}\mathrm{N}) = 
1/2^{+}$ leads in the weak coupling limit to a single $LS$ hypernuclear 
component $^{4}D_{1/2}$ for $^{15}_{\Lambda}\mathrm{N}$. Since the transition 
$^{15}_{\Lambda}\mathrm{N}({^{4}D_{1/2}})\to {^{15}\mathrm{O}({^{2}P_{1/2}})}$ 
requires spin-flip, it is forbidden for the dominant $\pi^-$-decay $s$-wave 
amplitude. In a more realistic calculation, for spin-parity $3/2^{+}$, the 
$^{15}\mathrm{O}$ ground-state main peak is expected to dominate over the 
secondary peak at about 6 MeV by a ratio close to 3:1, as shown in the lower 
part of Fig.~\ref{fig2a} by the red bars \cite{gal}. This is in rough 
agreement with the fitted gaussians shown by dashed lines in the upper part 
of Fig.~\ref{fig2a}, where the relative contribution of the $^{15}\mathrm{O}$ 
g.s. gaussian amounts to $(67 \pm 18)\%$. In contrast, for spin-parity 
$1/2^{+}$ the calculation in Ref.~\cite{gal} produces a ratio close to 1:1 
with respect to the $\sim 6$~MeV excitation, as shown by the blue bars 
in the lower part. Thus, the shape of the measured spectrum slightly favors  
$J^{\pi}(^{15}_{\Lambda}\mathrm{N}_{g.s.}) = 3/2^{+}$.\footnote{We note 
that the calculations in Refs.~\cite{motoba3,motoba1} suggest that neither 
the shape of the decay spectrum nor the total $\pi^-$ decay rate of 
$^{15}_{\Lambda}\mathrm{N}$  are sensitive to the assumed value of the 
ground-state spin. However, it has been shown recently \cite{gal} that these 
older results for $^{15}_{\Lambda}\mathrm{N}$ violate a model-independent 
sum rule and, therefore, are not used here further to suggest interpretations 
of the present experimental results.}

\section{MWD decay ratios and total decay rates} 
\label{p_tot} 

In general, due to the quite large errors affecting the spectra and to the 
lack of energy resolution at low values, the assignment of distinct MWD 
transitions to the daughter nucleus in our experimental spectra is somewhat 
tentative, except for the two-body component of $^{7}_{\Lambda}\mathrm{Li}$. 
However, by considering complementarily the total area of each spectrum it 
is possible to infer decay rates with a reasonable statistical significance. 

The branching ratios of the MWD reaction, 
$b_{\pi^{-}}=\Gamma_{\pi^{-}}/\Gamma_{tot}$, were evaluated for each 
hypernucleus as: 
\begin{equation} 
b_{\pi^{-}} = \frac{N_{\pi^{-} decay}}{N_{hyp}} 
\label{BR} 
\end{equation} 
where $N_{\pi^{-} decay}$ is the number of the $\pi^{-}$ MWD reactions 
and $N_{hyp}$ is the number of the produced hypernuclei. 

The number of MWD reactions was obtained from the counts in the momentum 
and the kinetic energy spectra after subtracting the residual background 
discussed in the previous section. In both cases, counts were considered 
up to the kinematical limit for a pure two-body decay, folded by our 
experimental resolution. 

\begin{table}[h] 
\begin{center} 
\begin{tabular}{|c|c|c|c|c|c|} 
\hline 
 &$b_{\pi^{-}}=\Gamma_{\pi^{-}}/\Gamma_{tot}$&$\Gamma_{tot}/\Gamma_{\Lambda}$& 
$\Gamma_{\pi^{-}}/\Gamma_{\Lambda}$ & previous data & theory \\  \hline 
$^{5}_{\Lambda}\mathrm{He}$ & $0.323 \pm 0.062^{+0.025}_{-0.020}$ & 
$1.03 \pm 0.08$ \cite{szym} & $0.332 \pm 0.069^{+0.026}_{-0.021}$ & 
$0.44 \pm 0.11$ \cite{szym} & 0.393 \cite{motoba4} \\ 
 & & $0.947 \pm 0.038$ \cite{kame} & $0.306 \pm 0.060^{+0.025}_{-0.020}$ & 
$0.340 \pm 0.016$ \cite{kame} & 0.305 \cite{gal}\\  \hline 
$^{7}_{\Lambda}\mathrm{Li}$ & $0.315 \pm 0.041^{+0.015}_{-0.012}$ & 1.12 
$\pm$ 0.12 & $0.353 \pm 0.059^{+0.017}_{-0.013}$ & & 0.304 \cite{motoba1} \\ 
 & & linear fit & & & (0.179) \\ 
 & & & & & 0.356 \cite{gal} \\ 
 & & & & & (0.176) \\ 
\hline 
$^{9}_{\Lambda}\mathrm{Be}$ & $0.154 \pm 0.040^{+0.011}_{-0.007}$ & 
$1.15 \pm 0.13$ & $0.178 \pm 0.050^{+0.013}_{-0.008}$ & & 0.172 \cite{motoba1} 
\\  & & linear fit & & & 0.186 \cite{gal}\\  \hline 
$^{11}_{\Lambda}\mathrm{B}$ & $0.199 \pm 0.039^{+0.041}_{-0.018}$ & 
$1.25 \pm 0.08$ \cite{park} & $0.249 \pm 0.051^{+0.051}_{-0.023}$ & 
$0.22 \pm 0.05$ \cite{mont} & 0.213 \cite{motoba1} \\ 
 & & $1.37 \pm 0.16$ \cite{grace} & & & (0.116) \cite{motoba2} \\  
 & & & & $0.23 \pm 0.06 \pm 0.03$ \cite{noumi} & 0.196 \cite{gal}\\ 
 & & & & $0.212 \pm 0.036 \pm 0.045$ \cite{sato} & (0.101) \cite{gal} \\ 
\hline 
$^{15}_{\Lambda}\mathrm{N}$ & $0.085 \pm 0.028^{+0.011}_{-0.010}$ & 
$1.26 \pm 0.18$ & $0.108 \pm 0.038^{+0.014}_{-0.013}$ & & 0.090 \cite{motoba1} 
\\  & & linear fit & & & (0.074) \cite{motoba_priv} \\ 
 & & & & & 0.080 \cite{gal}\\ 
 & & & & & (0.040) \cite{gal} \\ 
\hline 
\end{tabular} 
\caption{Branching ratios $b_{\pi^{-}}$, total hypernuclear weak decay rates 
$\Gamma_{tot}/\Gamma_{\Lambda}$ mostly from a linear fit in $A$, and total 
decay rates $\Gamma_{\pi^{-}}/\Gamma_{\Lambda}$ evaluated for charged MWD. 
Total decay rates are given in units of $\Gamma_{\Lambda}$. In the second 
and fourth columns the first quoted error is statistical, the second one is 
systematic. Comparison with previous measurements and theoretical predictions 
is reported. The calculated total rates are for ground state spin-parity 
$1/2^{+}$ for $^{7}_{\Lambda}\mathrm{Li}$, $5/2^{+}$ for 
$^{11}_{\Lambda}\mathrm{B}$ and $3/2^{+}$ for $^{15}_{\Lambda}\mathrm{N}$ 
(in brackets $3/2^{+}$ for $^{7}_{\Lambda}\mathrm{Li}$, 
$7/2^{+}$ for $^{11}_{\Lambda}\mathrm{B}$ and $1/2^{+}$ for 
$^{15}_{\Lambda}\mathrm{N}$).} 
\label{tab3} 
\end{center} 
\end{table} 

To evaluate the number of formed hypernuclei the area of the inclusive 
binding energy spectra was evaluated within the intervals reported in 
Table~\ref{tab1}, after subtraction of the $K^{-}np$ background and of 
the $K^{-}$ in flight decay contribution, as described above. 

The obtained values of the branching ratios, $b_{\pi^{-}}$, are reported in 
Table~\ref{tab3} with statistical and systematic errors. The latter ones are 
due to the different techniques used to evaluate the areas and the background 
in the inclusive spectra, while the systematic error due to the detection 
threshold of the apparatus has been estimated to be less than $2\%$ and has 
been neglected. A preliminary account of the results presented here can be 
found in \cite{panic08}. Total decay rates were calculated, using known 
$\Gamma_{tot}/\Gamma_{\Lambda}$ values or relying on a linear fit to the 
known values of all measured $\Lambda$-hypernuclei in the mass range 
$A=4-12$ \cite{sasao} as shown in Table~\ref{tab3}: 
\begin{equation} 
\Gamma_{tot}/\Gamma_{\Lambda}(A)=(0.990\pm 0.094)+(0.018\pm 
0.010) \cdot A,~~~~ \chi ^{2}/ndf = 5.317/6. 
\label{fit} 
\end{equation} 

A good agreement holds among the present results and previous measurements, 
when existing, and among the present results and theoretical calculations 
assuming ground state spins as listed in the caption. The calculated total 
decay rates for the other choice of ground state spins (in brackets) are 
substantially lower and disagree with the experimentally derived values. 
In particular, the total $\pi^-$ decay rate of $^{15}_{\Lambda}\mathrm{N}$ 
and its decay spectrum shape, as evaluated here, agree with calculations 
by Motoba {\it et al.} \cite{motoba1} and by Gal \cite{gal} assuming 
a ground-state spin-parity assignment $3/2^+$. These two calculations 
disagree for a $1/2^+$ spin-parity assignment, but as argued in 
Section~\ref{p_mwd} (footnote) the new calculation \cite{gal} for 
$^{15}_{\Lambda}\mathrm{N}$ corrects the older 
calculations \cite{motoba3,motoba1}. The calculation by Gal~\cite{gal} finds 
a significantly smaller total decay rate for $1/2^+$ spin-parity than for 
a $3/2^+$ spin-parity assignment, by $\sim 2\sigma$ lower than the rate 
evaluated by us as listed in the table. Based on this argument a spin-parity 
$1/2^+$ is excluded and a spin-parity $J^{\pi}(^{15}_{\Lambda}\mathrm{N}_{g.s.}) 
={ 3/2}^+$ assignment is made.

\section{Conclusions} 

We have reported a systematic study of MWD of $p$-shell $\Lambda$-hypernuclei 
by the FI-NUDA experiment, performing for the first time a magnetic analysis 
of spectra of $\pi^{-}$'s from MWD of $^{7}_{\Lambda}\mathrm{Li}$, 
$^{9}_{\Lambda}\mathrm{Be}$, $^{11}_{\Lambda}\mathrm{B}$ 
and $^{15}_{\Lambda}\mathrm{N}$. MWD decay rates 
$\Gamma_{\pi^{-}}/\Gamma_{\Lambda}$ have been evaluated and compared with 
previous measurements and theoretical calculations. The spin-parity 
assignments $J^{\pi}(^{7}_{\Lambda}\mathrm{Li}_{g.s.}) = {1/2}^+$ and 
$J^{\pi}(^{11}_{\Lambda}\mathrm{B}_{g.s.}) = {5/2}^+$ were confirmed 
and a new assignment, $J^{\pi}(^{15}_{\Lambda}\mathrm{N}_{g.s.}) = {3/2}^+$, 
was made based on the shape of the MWD spectra and the evaluated decay rates.

\section{Aknowledgements} 
The authors are grateful to Prof. Toshio Motoba for providing details on the 
calculations reported in Refs.~\cite{motoba2,motoba3,motoba1} and for the 
interest demonstrated in the FINUDA results.

\end{document}